\documentclass[prc,twocolumn]{revtex4-1}
\usepackage{graphicx}
\usepackage{dcolumn}
\usepackage{longtable}
\usepackage{amsmath}
\usepackage{amssymb}
\usepackage{natbib}

\begin{document}

\title{Bohr space in six dimensions}

\author{P. E. Georgoudis}
\email[]{georgoudis@inp.demokritos.gr}

\affiliation{Institute of Nuclear and Particle Physics, National Center for
Scientific Research "Demokritos", GR-15310 Aghia Paraskevi, Athens, Greece}

\begin{abstract}
A conformal factor in the Bohr model embeds Bohr space in six dimensions,
revealing the $O(6)$ symmetry and its contraction to the $E(5)$ at infinity.
Phenomenological consequences are discussed after the re-formulation
of the Bohr Hamiltonian in six dimensions on a five sphere.

\end{abstract}

\maketitle

\section{Introduction}

At subatomic scales matter is structured into particles,
but the interpretation of phenomena related to nuclear
collective motions in atomic nuclei, refers to a continuum
of nuclear matter with Riemannian geometry. Methods developed
in cosmology for the description of the spacetime continuum
at large scales, may also be used for the understanding of
such nuclear phenomena as well as the structural evolution
of atomic nuclei.

The comparison of the collective modes of motion of an
atomic nucleus with the oscillations of an irrotational
fluid \cite{Bohr2}, defines the reference to the continuum.
The fluid consists of nuclear matter \cite{Greiner},
which is achieved by the extension of the mass number $A$ at
infinity. Bohr space \cite{Bohr} is a five dimensional Riemannian
geometry for the continuum. Shapes of finite atomic nuclei,
axially symmetric, triaxial, and close to spherical
($\gamma$-unstable), can be represented by specific constraints
of Bohr space coordinates. But the initial reference to infinity,
which defines the continuum and thus the shape variables, is
absent in the Bohr geometry. A mass parameter obtained by
irrotational flow for the nuclear matter and the quantization
of the kinetic energy produce the Bohr Hamiltonian \cite{Bohr},
in which the comparison with the nuclear collective modes of
motion is implicit.

Nuclear collective modes of motion can also be analyzed in terms
of Interacting Bosons, paired valence nucleons of
angular momentum zero ($s$ boson) and two ($\bf{d}$ boson)
\cite{IBM}, which build the $U(6)$ symmetry group. Three different
chains of $U(6)$ subgroups define the dynamical symmetries of the
Interacting Boson Model the $U(5)$, $SU(3)$, and $O(6)$. In its
geometrical limit, obtained through the coherent states of $U(6)$
\cite{Dieperink}, the number of bosons extends to infinity which
also defines a continuum. Dynamical symmetries are thus translated
into phases of nuclear structure which accommodate spherical,
axially symmetric, and $\gamma$-unstable shapes, through energy
surfaces expressed in Bohr coordinates.

The relation between the continuum of Bohr geometry and that
of the geometrical limit of the IBM has been for many years
a matter of investigation. This is also the case in the study
of Quantum Phase Transitions \cite{CJC}, where the reference
to infinity in the context of the IBM is of primary importance
for the manifestation of these phenomena in the evolution of
structure of atomic nuclei. Their connection with shape transitions
in Bohr geometry and the possible symmetries which characterize
the transition, critical point symmetries, challenge again the
relationship between the geometrical limit of the IBM and Bohr
geometry.

Bohr geometry is five dimensional, while the geometrical limit of
the IBM \cite{Dieperink} is achieved from a geometrical constraint
on 5+1 coordinates which correspond to the coherent states of
the $U(6)$ group. The geometrical limit of the IBM succeeds to
offer an explicit relation between the nuclear collective modes
and the shape variables. An extra dimension in the Bohr geometry
can introduce the reference to infinity, as well as the emergence
of symmetries principally absent from the Bohr model.
Possible relations of the emergent symmetries with those of the IBM
should serve for the explicit comparison of the nuclear
collective motions with the oscillations of the nuclear fluid.

In the Bohr Hamiltonian, the absence of such an explicit
comparison could generate the well known phenomenological
deviations of the mass parameter. Apart from recent studies
\cite{Jolos}, they are also reflected in the discrepancy
with the experiment \cite{Scharff} of the relation between
the moments of inertia $\emph{J}$ and the intrinsic
quadrupole moment $\it{Q_{0}}$, $\emph{J}\propto \it {Q}^{2}_{0}$,
as it is predicted in the Bohr model through their
dependence with respect to the axial deformation $\beta$.
The hydrodynamic expression for the $\beta^{2}$-dependence
of the moments of inertia is certainly inconsistent
with the experiment. They exhibit a rather moderate
increase with respect to $\beta$ and are considerably
larger than the irrotational prediction \cite{Ring}.

It is the purpose of the present letter to propose a geometry
able to host the definition of the Bohr space at infinity.
Success relies on the embedding of Bohr space in six dimensions
as a five sphere. A conformal factor compares the five sphere
with the initial Bohr space, which is revealed as the boundary at
infinity. The symmetry group of the five sphere is the $O(6)$,
its relation with the correspondent limit of the IBM is discussed,
and contracts to the $E(5)$ symmetry \cite{Iachello} at infinity.

Re-formulation of the Bohr Hamiltonian in six dimensions, on the
five sphere and on the projected plane, introduces the
phenomenological consequences of the proposed geometry. They
begin from the alteration of the dependence of the moments of
inertia with respect to the deformation, or the appearance of
a deformation-dependent mass parameter. Forms of the $\beta$-part
of the moments of inertia depend on the symmetry region, implying
symmetry-dependent rules for their relation with the intrinsic
quadrupole moment. Last but not least, the emergent presence
of $E(5)$ opens new possibilities for its manifestation in
atomic nuclei.
\section{Embedding in six dimensions}
Bohr space can be defined by the element of length
 \begin{equation}
 ds^{2}=g_{ij}dq^{i}dq^{j},
 \end{equation}
with $q^{1}=\beta$, $q^{2}=\gamma$, $q^{3}=\Theta$, $q^{4}=\Phi$,
$q^{5}=\Psi$. These coordinates span a five dimensional space $R^{5}$.
It is defined \cite{Rowe} by the tensor product of the radial line
$R_{+}$, which represents the totality of the $\beta$ values, and the
unit four sphere $S_{4}$, which represents the totality of the values
of the angle $\gamma$ and the three Euler angles $\Theta, \Phi,\psi$.
Namely $R^{5}\sim R_{+}\times S_{4}$. In what follows, the notation
$(\beta,\Omega_{4})$ for the Bohr coordinates $q^{i}$ will be useful,
with $\Omega_{4}$ representing the four angles.

If $d\Omega_{4}$ is a line element on the unit four sphere $S_{4}$,
then a consistent form of an $R^{5}$ line element with the definition
of $R^{5}\sim R_{+}\times S_{4}$ is
\begin{equation}\label{Bohr}
ds^{2}=d\beta^{2}+\beta^{2} d\Omega^{2}_{4}.
\end{equation}
As in \cite{DDM}, a conformal transformation can take place
in the Bohr metric
 \begin{equation}
 g_{ij}\rightarrow \tilde{g}_{ij}=\frac{1}{(1+a\beta^{2})^{2}}g_{ij},
 \end{equation}
dependent on the variable $\beta$ and a parameter $a$.
The line element $ds$ affected by the conformal factor
gives a $dl$ with
\begin{equation}\label{conf}
{dl^{2}}=\frac{1}{(1+a\beta^{2})^{2}}(d\beta^{2}+\beta^{2}
 d\Omega^{2}_{4}).
\end{equation}
Such forms of line elements exist in cosmological models, like the
Robertson-Walker, for the description of conformally flat spaces.
They are derived by the projection of a hypersphere to the plane,
as can be found in a typical textbook in general relativity \cite{Landau}.
As a formal correspondence with the geometry of RW, pick a system
of coordinates of five angles $(\chi,\Omega_{4})$, with $\chi=\chi(\beta)$,
and $\Omega_{4}$ the same as in Bohr space. The
totality of the values of the five angles $(\chi,\Omega_{4})$
at a common radius $R$, define the five sphere $S_{5}$ with
line element
\begin{equation}\label{sph}
 dl^{2}=R^{2}(d\chi^{2}+\sin^{2}\chi d\Omega^{2}_{4}).
\end{equation}
An appropriate transformation between the angle
$\chi$ and the radial variable $\beta$, so as to produce the
line element of Eq. (\ref{conf}) is obtained by standard
calculations of the RW \cite{Landau} geometries.
Choosing
\begin{equation}
R=\frac{1}{\sqrt{4a}}, \ \ \sin\chi=\tilde{\beta}\sqrt{4a},
\end{equation}
then
\begin{equation}\label{rw}
 dl^{2}=\frac{d\tilde{\beta^{2}}}{1-4a\tilde{\beta}^{2}}+
 \tilde{\beta}^{2}d\Omega^{2}_{4}.
 \end{equation}
 Definition of the radial coordinate $\beta$ from
\begin{equation}\label{vr}
  \tilde{\beta}=\frac{\beta}{1+a\beta^{2}},
\end{equation}
gives the line element (\ref{conf}).
Examination of the $\tilde{\beta}$ values in Eq. (\ref{vr})
for limit values of $\beta$, illustrates the geometry. For
$\beta=0$, $\tilde{\beta}$ is zero and for $\beta \rightarrow \infty$,
$\tilde{\beta}$ goes to zero. The beginning and the end of the
$\tilde{\beta}$ for the totality of $\beta$ values is the same point.
Angle $\chi$ geometrizes this circle feature of the $\tilde{\beta}$,
with the interval of its values to be $0\leq\chi\leq\pi$.

The $S_{5}$ lives in a $5+1$-dimensional Euclidean space.
In coordinates $q^{u}$, $_{(u=0,\ldots,5)}$, the $S_{5}$
of radius $\frac{1}{\sqrt{4a}}$ defines a boundary in
six dimensions, represented by the equation
\begin{equation}\label{bound}
(q^{0})^{2}+ \sum^{5}_{u=1}q^{u}q^{u}=\frac{1}{4a}.
\end{equation}
Stereographic projection from the north pole of the $S_{5}$
on to the tangent plane is equivalent with the conformal
transformation of the Bohr line element. The tangent
plane is the $R^{5}$, spanned by the quadrupole
degree of freedom $q^{m}$, $_{(m=2,1,0,-1,-2)}$.
Standard relationships of stereographic projection
\cite{Robertson} give
\begin{equation} \label{stereo}
q^{u}=\frac{q^{m}}{1+a( \hat{q}\cdot\hat{q})},\quad q^{0}=\frac{1-a(
\hat{q}\cdot\hat{q})}{\sqrt{4a}(1+a( \hat{q}\cdot\hat{q}))},
\end{equation}
with the tangent plane to be located at $q^{0}=-\frac{1}{\sqrt{4a}}$.
Now, $\tilde{\beta}$ is
revealed as the projected radial variable from $S_{5}$
to the tangent plane
\begin{equation}
{\tilde{\beta}}^{2}=\sum^{5}_{u=1}q^{u}q^{u}=\frac{1}{(1+a(
\hat{q}\cdot\hat{q}))^{2}}\sum_{m}q^{2}_{m}.
\end{equation}
A part of the tangent plane is characterized by the
$\tilde{\beta}$ variable. This is the projected plane,
defined by (\ref{rw}), which is compared with the Bohr
space through the conformal factor. For $a=0$, the boundary
of $S_{5}$ extends to infinity, where the Bohr line element
$ds$ is obtained.

Any well defined boundary in six dimensions can be characterized as
a Bohr hypersurface. This reflects the possibility of the
existence of Bohr type Hamiltonians in these hypersurfaces.
The original Bohr hamiltonian belongs to the boundary at
infinity. On the other hand, the symmetry group of the totality
of transformations of the $S_{5}$ into itself is the $O(6)$.
\section{O(6) and the IBM}
Let us construct the generators of $O(6)$ in terms of
$q^{u}$ and their conjugate momentum generators $\partial_{u}$.
They are obtained as the special case of $n=5$ for
the $O(n+1)$ \cite{Gilmore}. With $u,k\neq0$, $O(6)$
generators are
\begin{equation}\label{osix}
Q_{uk}=q^{u}\partial_{k}-q^{k}\partial_{u}, \ \ Q_{u0}=q^{u}
\partial_{0}-q^{0}\partial_{u}.
\end{equation}
In \cite{Castanos} the IBM is considered as a six dimensional
harmonic oscillator, with the $s$ and $\bf{d}$ bosons to be
expressed in terms of six variables, and their canonical conjugates,
which we write as $(q^{0},q^{m})$. The $O(6)$ limit reveals a structure
like (\ref{osix}) in a geometry of one radial variable $b$, and five
angles $(\delta,\Omega_{4})$, with
\begin{equation}
q^{0}=b\cos\delta, \quad \beta=b\sin\delta.
\end{equation}
The embedding of Bohr space in six dimensions is characterized
by the variables
\begin{equation}\label{C}
q^{0}=\frac{1}{\sqrt{4a}}\cos\chi, \quad \tilde{\beta}=
\frac{1}{\sqrt{4a}}\sin\chi.
\end{equation}
For $b=\frac{1}{\sqrt{4a}}$, and $\tilde{\beta}$ instead of $\beta$,
$\delta=\chi$. Therefore, the $S_{5}$ is that of the $O(6)$ limit
of the IBM, if in \cite{Castanos} $\tilde{\beta}$ replaces $\beta$
which induces its stereographic projection to the tangent plane.
The number operator \cite{Castanos} is
\begin{equation}\label{number}
\hat{N}=\frac{1}{2}\left(-\frac{1}{b^{5}}\frac{\partial}
{\partial_{b}}b^{5}\frac{\partial}{\partial_{b}}+
\frac{1}{b^{2}}\emph{L}^{2}+b^{2}\right)-3,
\end{equation}
with $\emph{L}^{2}$ the second order Casimir of $O(6)$. Replacing
$\beta$ with $\tilde{\beta}$ in \cite{Castanos}
\begin{equation}
\emph{L}^{2}=-\left(\frac{1}{\sin^{4}\chi}\frac{\partial}
{\partial_{\chi}}\sin^{4}\chi
 \frac{\partial}{\partial_{\chi}}-\frac{1}{\sin^{2}\chi}
\Lambda^{2}\right).
\end{equation}
The choice of $b=\frac{1}{\sqrt{4a}}$ is a constraint in the
number operator (\ref{number}), its radial part is canceled.
For $a=0$, $N$ goes to infinity.

\section{Re-formulation of the Bohr Hamiltonian}

The re-formulation of the Bohr Hamiltonian on
the $S_{5}$ can be obtained by the laplacian of the
line element (\ref{sph}), assuming a mass parameter
$B$ as in the initial Bohr hamiltonian \cite{Bohr}.
The result is
\begin{equation}\label{Bo6}
H=-\frac{\hbar^{2}}{2B}\left(\frac{4a}{\sin^{4}\chi}\frac{\partial}
{\partial_{\chi}}\sin^{4}\chi \frac{\partial}{\partial_{\chi}}
-\frac{4a}{\sin^{2}\chi}
\Lambda^{2}\right).
\end{equation}
$\Lambda^{2}$ is the second order Casimir operator of the $SO(5)$.
The projection from $S_{5}$ generates a distinct Hamiltonian, with
the presence of the conformal factor. It is obtained by the laplacian
of the line element (\ref{conf}), which gives
\begin{equation}\label{proj}
H=-\frac{\hbar^{2}}{2B}\left[\frac{(1+a\beta^{2})^{5}}{\beta^{4}}
\frac{\partial}{\partial_{\beta}}\frac{\beta^{4}}
{(1+a\beta^{2})^{3}}\frac{\partial}{\partial_{\beta}}-\frac{
(1+a\beta^{2})^{2}}{\beta^{2}}
\Lambda^{2}\right].
\end{equation}
Apart from $B$, the former Hamiltonian is the second term
of the Number operator (\ref{number}), which contains the
second order Casimir of the $O(6)$ of the IBM. Again, the
two expressions coincide if in \cite{Castanos} $\tilde{\beta}$
replaces $\beta$.

The latter Hamiltonian (\ref{proj}), lives on the projected
plane from the $S_{5}$. The conformal factor affects, apart
from the whole radial part, only the coefficient
of $\Lambda^{2}$. Now the dependence
of the moments of inertia with respect to $\beta$ is
$\frac{\beta^{2}}{(1+a\beta^{2})^{2}}$. The same result is
obtained in Bohr's calculation \cite{Bohr}
\begin{equation}\label{moi}
\emph{J}_{k}=B\sum_{uu^{\prime}}q_{u}q_{u^{\prime}}
(M^{2}_{k})_{uu^{\prime}}=4B\frac{\beta^{2}}
{(1+a\beta^{2})^2}sin^{2}\left(\gamma-k\frac{2\pi}{3}\right),
\end{equation}
using the coordinates $q^{u}$,$_{(u\neq0)}$, of Eq. (\ref{stereo}).
$M$ is the angular momentum. The angles $\Omega_{4}$ were not
affected from the embedding and the Wigner matrices are not
influenced by the projection, as of the invariance of the
$O(5)\supset O(3)$.

The effects of the projection from the $S_{5}$,
or of the $O(6)$ symmetry, can appear in
a deformation-dependent mass parameter in the
initial Bohr space without the extra dimension.
This is indicated by the last expression (\ref{moi})
where a mass parameter $B(\beta)=\frac{B}{(1+a\beta^{2})^{2}}$
and the Bohr coordinates $q^{m}$, produce the same result. A Bohr
Hamiltonian with this deformation-dependent mass parameter
has been proposed in \cite{DDM}.

The reduction of the moments of inertia in the present
approach implies that their relation with the
intrinsic quadrupole moment $Q_{0}$ is symmetry-dependent.
The original Bohr Hamiltonian in $R^{5}$ gives moments of inertia
$\emph{J}_{k}\propto \beta^{2}$, while that of (\ref{proj})
gives $\emph{J}_{k}\propto \beta^{2}/ (1+a\beta^{2})^{2}$.
Projection from $S_{5}$ is an $O(6)$ fingerprint, and signifies
a new symmetry region than that of the original Bohr Hamiltonian
in $R^{5}$. With the usual quadrupole operator of the Bohr model
\cite{Caprio} $Q_{0} \propto \beta$. The new rule for the relation
of the moments of inertia with the intrinsic quadrupole moment is
$\emph{J}_{k}\propto Q_{0}^{2}/(1+aQ_{0}^{2})^{2}$.
However the exact rule needs the consideration of the $\beta$-part
of the quadrupole operator in the new symmetry region. This is beyond
the purposes of the present work, it should be done in agreement
with the geometrical limit of the IBM.
\section{E(5)}

To which region of the IBM shall the Hamiltonian (\ref{proj})
be located? It should not be the exact
$O(6)$ limit, as it does not live on the five sphere but on the
projected plane. On the other hand, the symmetries revealed by
the projection of $S_{5}$ to $R^{5}$, which involves a dimensional
reduction from six to five dimensions, can be studied with the use
of the group contractions.

At the boundary of $S_{5}$ at infinity,
$O(6)$ contracts \cite{Gilmore} to the Euclidean group
in five dimensions $E(5)$.
\begin{figure}
 \includegraphics[height=7.1 cm]{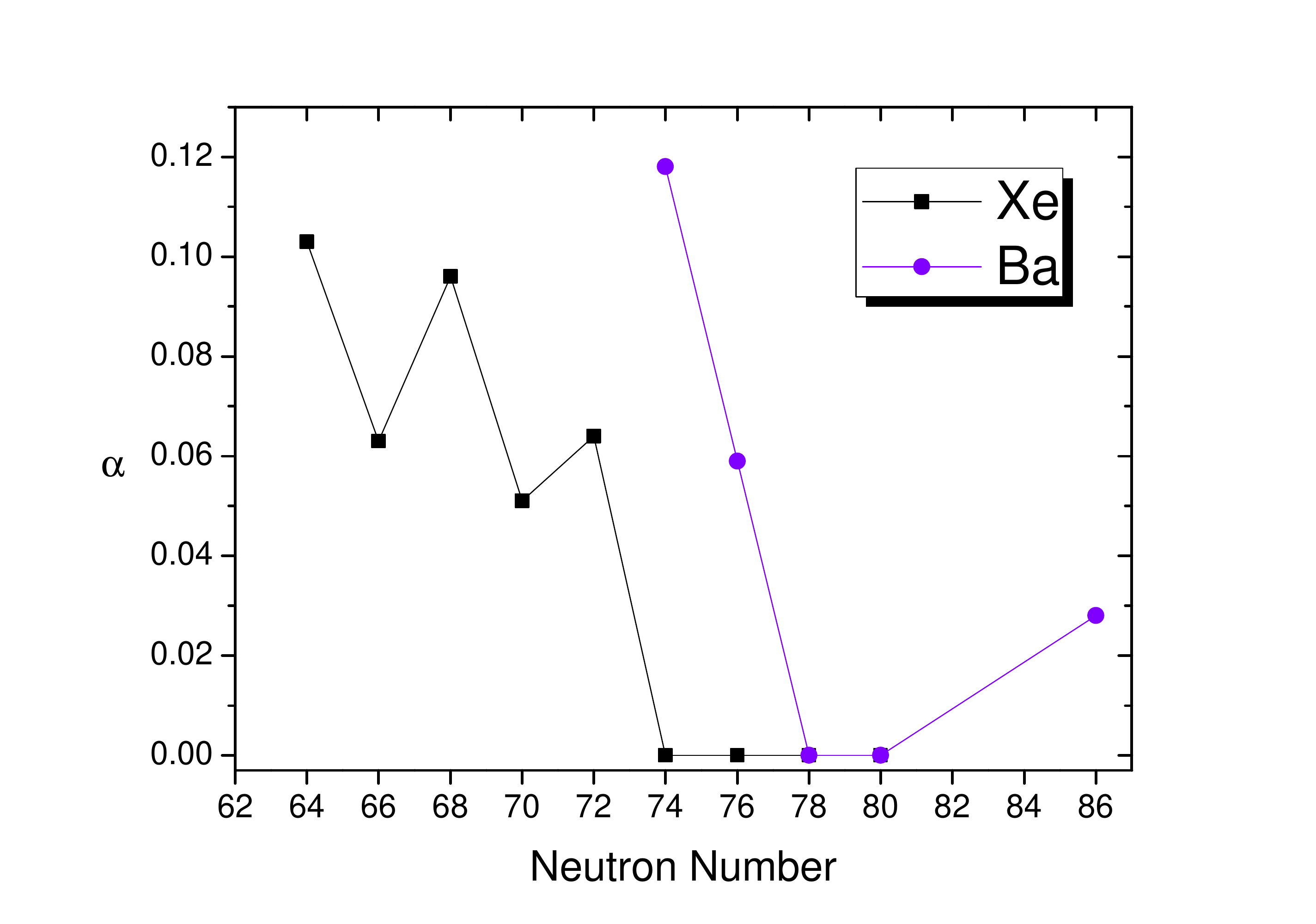}
 \caption{\label{fg1} Values of $a$ for Xe and Ba isotopes,
 data taken from \cite{DDM}.}
 \end{figure}
Contraction process \cite{Gilmore} examines the $Q_{u0}$ (\ref{osix})
at the north pole of $S_{5}$, where $q^{0}=\frac{1}{\sqrt{4a}}$
and the rest $q^{u}=0$,
\begin{equation}
Q_{uo}=-\frac{1}{\sqrt{4a}}\partial_{u}.
\end{equation}
This expression is singular for $a=0$.
Define the operator
\begin{equation}
\tilde{Q}_{u0}=\sqrt{4a}(q^{u}\partial_{0}-q^{0}\partial_{u}).
\end{equation}
Now at the north pole and for $a=0$,
$\tilde{Q}_{u0}=-\partial_{m}$. The set of
generators $\{Q_{uk},\tilde{Q}_{u0}\}$ contracts to the
set $\{{Q_{mn},\partial_{m}}\}$ for $a=0$. The latter set
generates the $E(5)$ symmetry group \cite{Caprio},
which was proposed in \cite{Iachello} to characterize the
critical point for a second order shape-phase transition
from spherical to $\gamma$-unstable nuclei. It corresponds
to the infinite square well potential for the $\beta$
variable, in the initial Bohr Hamiltonian. The Hamiltonian
(\ref{proj}) for $a=0$ coincides with that of $E(5)$
\cite{Iachello}. Now, the geometrical limit of IBM \cite{IBM}
in its $U(5)-O(6)$ transitional region hosts a point for
a second order phase transition. The presence of $E(5)$
for $a=0$, the $O(5)$ invariance during the projection,
as well as the argument that the Hamiltonian (\ref{proj})
should not correspond to an exact $O(6)$ limit, imply its
adaptation to the $U(5)-O(6)$ transitional region of the IBM.

Predictions of $E(5)$ manifestations in atomic nuclei which
lean on the spectrum of the initial Bohr Hamiltonian,
are rather incomplete. The absence of the conformal factor
in the initial Bohr Hamiltonian reflects the absence of
$O(6)$ fingerprints. Parameter $a$ controls the presence
of the conformal factor. In \cite{DDM}, $a$-values are
obtained by their rms fitting to the experimental
data for the energy spectrum of a plethora of $\gamma$-unstable
nuclei. A close examination of its available values draws
attention to the nuclei for which $a=0$. Fig. \ref{fg1}
displays the a-values for the Xe and Ba isotopes.
Parameter $a$ is not zero for a single nucleus in the
series of Xe, or in that of the Ba isotopes. $a=0$
corresponds to the series of $^{128}$~Xe, $^{130}$~Xe,
$^{132}$~Xe, $^{134}$~Xe as well as in those of $^{134}$~Ba
and $^{136}$~Ba. This remark should not be received as a
proposal for $E(5)$ candidates. In \cite{DDM} for $a=0$
the Hamiltonian of $E(5)$ \cite{Iachello} is not obtained,
Davidson term $\beta^{2}$ survives which nevertheless has
been proposed to characterize the transitional region
\cite{Rowe}. The present approach implies that the
manifestation of $E(5)$ in atomic nuclei, needs additional
experimental measures than those obtained by the initial
Bohr Hamiltonian. A geometrical limit of the IBM containing
the parameter $a$ should illustrate its appearance in nuclear
structure.

\section{Conclusions}
Beginning from a formal correspondence of
Bohr space with the conformal factor and the cosmological
Robertson-Walker geometry, Bohr model is embedded in six dimensions.
The original Bohr space is located at the limit of infinite radius
of a five sphere. Bohr Hamiltonian is re-formulated on the five
sphere, which has the $O(6)$ symmetry group. Its relation with
the $O(6)$ limit of the IBM indicates that the infinity of the
five sphere corresponds to the limit of infinite number of bosons
in the IBM.

The five sphere with finite radius corresponds to a constraint
on the Number operator of the IBM. Bohr space with the conformal
factor coincides with the stereographic projection of the five
sphere on to the tangent plane. The re-formulation of the Bohr
Hamiltonian on the projected plane is adapted to the $U(5)-O(6)$
region of the IBM. Its consequences are:

(i) The appearance of a deformation-dependent mass parameter
    in the original Bohr model.

(ii) Equivalently, in the projected plane
     the moments of inertia exhibit a different dependence
     with respect to the deformation, than in the original
     Bohr model. This implies the existence of symmetry-dependent
     rules for the relation of the moments of inertia
     with the intrinsic quadrupole moment.

(iii) In the limit of infinite radius of the five sphere,
      $E(5)$ Hamiltonian emerges. The new way of $E(5)$
      generation challenges a discussion about its possible
      manifestations in atomic nuclei.

\begin{acknowledgments}
I wish to thank D. Bonatsos for a critical reading
of the manuscript and useful discussions.
Also, I am thankful to P. Van Isacker for enlightening
discussions, especially those concerning
the re-formulation of the Bohr Hamiltonian in six
dimensions on a five sphere.
\end{acknowledgments}






\begin{thebibliography}{99}

\bibitem{Bohr2}
A. Bohr and B. Mottelson, Dan. Mat. Fys. Medd.
{\bf 30}, (1955), 1.

\bibitem{Greiner}
W. Greiner and J.A. Maruhn, Nuclear Models
Springer-Verlag, Berlin, 1996.

\bibitem{Bohr}
A. Bohr, Dan. Mat. Fys. Medd. {\bf 26}, (1952), 14.

\bibitem{IBM}
F. Iachello and A. Arima, The Interacting Boson Model
Cambridge University Press, Cambridge, 1987.

\bibitem{Dieperink}
A.E.L. Dieperink, O. Scholten and F. Iachello, Phys.
Rev. Lett. {\bf 44}, (1980), 1747.
J. N. Ginocchio and M. W. Kirson, Phys. Rev. Lett. {\bf 44}, (1980),
1744.

\bibitem{CJC}
P. Cejnar, J. Jolie and R. F. Casten, Rev. Mod.
Phys., {\bf 82}, (2010).

\bibitem{Jolos}
R. V. Jolos and P. von Brentano, Phys. Rev. C {\bf 79},
(2009), 044310.

\bibitem{Scharff}
G. Scharff-Goldhaber, C.B. Dover and A.L. Goodman,
Ann. Rev. Nucl. Sci. 26, (1976) 239-317.

\bibitem{Ring}
P. Ring and P. Schuck, The Nuclear Many- Body Problem, Springer-Verlag, Berlin, 1980.

\bibitem{Iachello}
F. Iachello, Phys. Rev. Lett. {\bf 85}, (2000), 3580.

\bibitem{DDM}
D. Bonatsos, P.E. Georgoudis, D.Lenis, N.Minkov,
and C.Quesne, Phys. Rev. C \textbf{83},
(2011), 044321.

\bibitem{Rowe}
D.J. Rowe, T.A. Welsh and M.A. Caprio, Phys. Rev.
C {\bf79}, (2009), 054304. See also David J.Rowe
and John L.Wood, Fundamentals of Nuclear Models
World scientific, London, 2009.


\bibitem{Landau}
L.D. Landau and E.M. Lifshitz, The classical
theory of fields, Butterworth-Heinemann, Oxford,
1999.

\bibitem{Robertson}
H.P. Robertson, Rev. Mod. Phys. {\bf 5}, 62, (1933).

\bibitem{Gilmore}
Robert Gilmore, Lie Groups, Lie Algebras
and Some of their Applications, John Wiley
and Sons, New York, 1974.


\bibitem{Castanos}
O. Castanos, E. Chacon, A. Frank and M. Moshinsky, J. Math. Phys. {\bf 20}, (1979), 1.

\bibitem{Caprio}
M.A. Caprio and F. Iachello, Nucl. Phys.
A{\bf781}, (2007), 26.

\end{thebibliography}



\end{document}